\documentclass[11pt]{article}

\usepackage{moriond,epsfig}
\usepackage{amsmath}
\usepackage{amssymb}
\usepackage{euscript}
\usepackage{bbm}
\usepackage{epsfig}

\def\be{\begin{equation}}
\def\ee{\end{equation}}
\def\bea{\begin{eqnarray}}
\def\eea{\end{eqnarray}}
\def\bse{\begin{subequations}}
\def\ese{\end{subequations}}
\def\bc{\begin{cases}}
\def\ec{\end{cases}}

\newcommand{\Keu}{\EuScript{K}}

\newcommand{\Kbm}{\mathbbm{K}}

\newcommand{\veps}{\varepsilon}

\begin{document}

\vspace*{4cm}
\title{EVOLUTION EQUATIONS IN QCD AND QED}
\author{ M. SLAWINSKA}
\address{Institute of Nuclear Physics, Polish Academy of Sciences,\\
	ul.\ Radzikowskiego 152, 31-342 Cracow, Poland.}

\maketitle\abstracts{ 
Evolution equations of YFS and DGLAP types in leading order are considered. They are compared in terms of mathematical properties and solutions. In particular, it is discussed how the properties of evolution kernels affect solutions. Finally, comparison of solutions obtained numerically are presented.}

\section{Introduction}

Generic QCD evolution equation covering several types of evolution reads:
\begin{equation}
\label{eq:genevoleq}
  \partial_t D(t,x)
= D(t, \cdot)\otimes  \Keu(\cdot)(x),  
\end{equation}
where the change of a structure function $D$ with a process scale $t$ is given by its convolution with the evolution kernel $\Keu$. I will set the boundary condition $D(0,x)=\delta(1-x)$.

Equations of the type (\ref{eq:genevoleq}) appear very often in high energy physics. For the purpose of this work, we will consider two simplified cases, originating from QED and QCD radiative processes. They were chosen due to their mathematical similarities to demonstrate how properties of evolution kernels influence their solutions.

Let us consider the evolution equation with an additive convolution rule:
\begin{equation}\label{dyfs_0}
\partial_t D_{YFS}(t,x)
= \int_0^1 dy du\; \delta(x-y-u)\Keu_{YFS}(y) D_{YFS}(t,u)
 = \int_0^x du\; \Keu_{YFS}(x-u) D_{YFS}(t,u)
\end{equation}
with the kernel given by:
\begin{equation}
\Keu_{YFS}(u, x)\label{Keu_YFS}
= \Theta(u-x)\Theta(1-u)\frac{1}{u-x} - \delta(u-x)\int_x^{1+x}\frac{du}{u-x}.
\end{equation}

The distribution $D_{YFS}$ has been calculated by Yennie, Frautschi, and Suura ~\cite{yfs:1961} in the context of exponentiation infrared singularities in QED\footnote{Precisely, we consider distributions obtained from the original one with the substitution $x\rightarrow 1-x$ and $u\rightarrow 1-u$.The definition of $\Keu_{YFS}$ follows from the famous photon spectrum $\Kbm(x)= \frac{1}{x} - \delta(x)\int_0^1\frac{dv}{v}$}.

Another example is the evolution equation of DGLAP type~\cite{DGLAP}, describing evolution of gluonic momentum distribution in the leading order:
\begin{equation}\label{DGLAP}
\partial_t D(t, x) = \int_0^1 dz du\; \delta(x-uz)\Keu_{DGLAP}(u) D(t,z) \equiv\int_x^1 \frac{dz}{z} \Keu_{DGLAP}\left(\frac{x}{z}\right) D(t,z),
\end{equation}
The function $D(x)$ is by some authors denoted as $xD(x)$~\cite{Ellis:1991qj}. The QCD sum rule requires
\begin{equation}\label{sumrule}
\int_0^1 dx \; D(t, x)=1.
\end{equation}
The above can be fulfilled if
\begin{subequations}
\begin{equation}
\int_0^1 dx \; D(0, x)=1, \\
\end{equation}
\begin{equation}
\partial_t \int_0^1 dx \; D(t, x) = 0.
\end{equation}
\end{subequations}
The former condition is assured by initial condition, the latter may be obtained from (\ref{DGLAP}):
\begin{equation}
\begin{split}
\partial_t \int_0^1 dx D(t|x)&= \int_0^1 dx \int_0^1 dz \int_0^1du \delta(x-uz)\Keu_{DGLAP}(z) D(t,u) \\
&= \int_0^1 dz\Keu_{DGLAP}(z) \int_0^1 du  D(t,u)
\end{split}
\end{equation}
and is fulfilled if
\begin{equation}\label{Pnorm}
\int_0^1 dz\Keu_{DGLAP}(z) = 0.
\end{equation}
The condition \eqref{Pnorm} is assured by the definition $\Keu_{DGLAP}=\left(\frac{1}{1-x}\right)_+$

\section{Comparison}

Evolution equations (\ref{dyfs_0}) and (\ref{DGLAP}) are governed by similar kernels, with major difference in the convolution type. To compare them  we will introduce regularized kernels and transform additive evolution into  multiplicative. That form would enable us to see that the difference between them lies in the details of kernels definitions. 

In order to find out interrelations between both evolutions we will work using regularized kernels. Both kernels can be expressed as a sum of regular (``$\Theta$'') and singular (``$\delta$'') parts

\be\label{genreg}
\Keu(x) \equiv \Keu^{\Theta}(x)\Theta(1-x-\veps) - \Keu^{\delta}(x)\delta(1-x-\veps).
\ee
Parameter $\veps \ll 1$ plays the role of a constant ($x$-independent) infrared cutoff.

We will write both kernels in the form they are used in evolution equations (\ref{dyfs_0}) and (\ref{DGLAP}), as functions of two variables.

In the YFS case:
\begin{equation}\label{reg_K}
\Keu_{YFS}(u,x)  
= \frac{\Theta(u-1)}{u-x}\Theta(u-x- \veps) - \delta(u-x-\veps)\ln\frac{1}{\veps}.
\end{equation}
The DGLAP kernel $\Keu_{DGLAP}$ is usually regularized in the $(x, z)$ plane as follows:
\begin{equation}\label{reg_P}
\Keu_{DGLAP}(z) =\frac{1}{1-z}\Theta(1-z - \veps) - \delta(z-1-\veps)\int_0^{1-\veps} dz \frac{1}{1-z}.
\end{equation}

The above notation stresses that the kernels enter their evolution equations regularized  in different variables, according to their convolution rules. In the YFS case $\veps$ is defined in ``additive'' variables $(x, u)$ and assures $x-u > 0$. The DGLAP kernel is usually regularized ``multiplicatively'', such that $\frac{x}{z}<1$. 

Simple transformation of  \eqref{DGLAP} leads to an evolution equation resembling the YFS additive convolution \eqref{dyfs_0}:

\begin{equation}\label{DGLAP->YFS}
\begin{split}
\partial_t D(t,x) &= \int_{x}^{1}\left[ \frac{\Theta(1- z -\veps)}{1-z}D\left(t,\frac{x}{z}\right)\frac{1}{z}-\delta(z-1)\ln\frac{1}{\veps} D\left(t|\frac{x}{z}\right)\frac{1}{z}\right]dz \\
&= \int_{x }^{1}\left[ \frac{\Theta(1- x/u -\veps)}{u-x}D(t,u)-\delta(x-u)\ln\frac{1}{\veps} D(t,x)\right]du.\\
\end{split}
\end{equation}

$\Keu_{DGLAP}$ in (\ref{DGLAP->YFS}), however is regularized in $(x, z)$ variables, which means that the cutoff is $u$-dependent. This regularization is given by $\Theta(1-x/u - \veps)$, which transforms into $u \geq \frac{x}{1+\veps}$. From (\ref{DGLAP->YFS}) it is already transparent that  different regularization of evolution kernels led to differences in their ``$\Theta$'' parts.

Since $\veps$ is just a parameter set 0 at the end, one can go a step further and regularize DGLAP kernel on the $(x, u)$ plane with constant parameter $\veps'$: $u-x > \veps'$ and compare both kernels in that language. In that case, $\veps$ in  \eqref{DGLAP->YFS} is $x$-dependent and should be expressed by $x$ and $\veps'$. It could be done as follows:
Let $\veps$ be a (so far not specified) function of $x$ and $\veps'$~\footnote{Redefinition $\veps \rightarrow \veps(x)$ is a legal operation, as $x$ is just a parameter of the integrand \eqref{DGLAP->YFS}.}. One can rewrite the last line of \eqref{DGLAP->YFS} in this new regularization:

\begin{equation}\label{DGLAP_reg_xu}
\begin{split}
\partial_t D(t,x)
&= \int_{x }^{1}\left[ \frac{\Theta\left(u- x -\frac{\veps(x, \veps')}{1-\veps(x, \veps')}x\right)}{u-x} D(t,u)-\delta(x-u)\ln\frac{1}{\veps(x, \veps')} D(t,x)\right]du,
\end{split}
\end{equation}

from which it follows that the requirement of constant cutoff $\veps'$ leads to a formula
\begin{equation}\label{vepsveps'}
\veps' = \frac{\veps(x, \veps')}{1-\veps(x, \veps')}x
\end{equation}
relating both cutoffs. From \eqref{vepsveps'} we obtain $\veps(x, \veps') = \frac{\veps'}{x+\veps'} \approx \frac{\veps'}{x}$ so that DGLAP evolution kernel regularized on the $(x, u)$ plane by constant $\veps'$ has the form:

\begin{equation}\label{reg2_P}
\Keu_{DGLAP}(x, u) = \Theta(u-x- \veps')\frac{1}{u-x} - \delta(u-x)\ln\frac{x}{\veps'}. 
\end{equation}

Now it is clear that both kernels, although regularized identically and having equal ``$\Theta$'' parts differ in their singular parts and are therefore different distributions. 

\section{Normalization properties}

Another remark concerns normalization of both kernels. It is an intrinsic property of kernels, independent of regularization techniques and following from definitions. $\Keu_{DGLAP}$ is normalized according to \eqref{Pnorm}, whereas  $\Keu_{YFS}$ inherits its normalization from $\Kbm(v)$. $\int_0^1 \Kbm(v)dv  = \int_{u-1}^{u}\Kbm(u,x)dx = 0$, whereas the integration range for $\Keu_{YFS}(u,x)$ is $[0, u]$. Therefore 
\begin{equation}
\int_0^u\Keu_{YFS}(u,x)dx =  \int_0^u \Kbm(v)dv = a\ln u \leq 0.
\end{equation}

As  $\Keu_{YFS}(u,x)$ is not normalized according to evolution equation, the normalization of $D_{YFS}$ is time-dependent (and decreasing). Normalization properties of $\Keu_{DGLAP}$ are imposed by (\ref{sumrule}), so $D_{DGLAP}$ is normalized by construction.

\section{Comparison of solutions}

\begin{figure}\label{YFS_DGLAP}
\begin{centering}
\epsfig{file=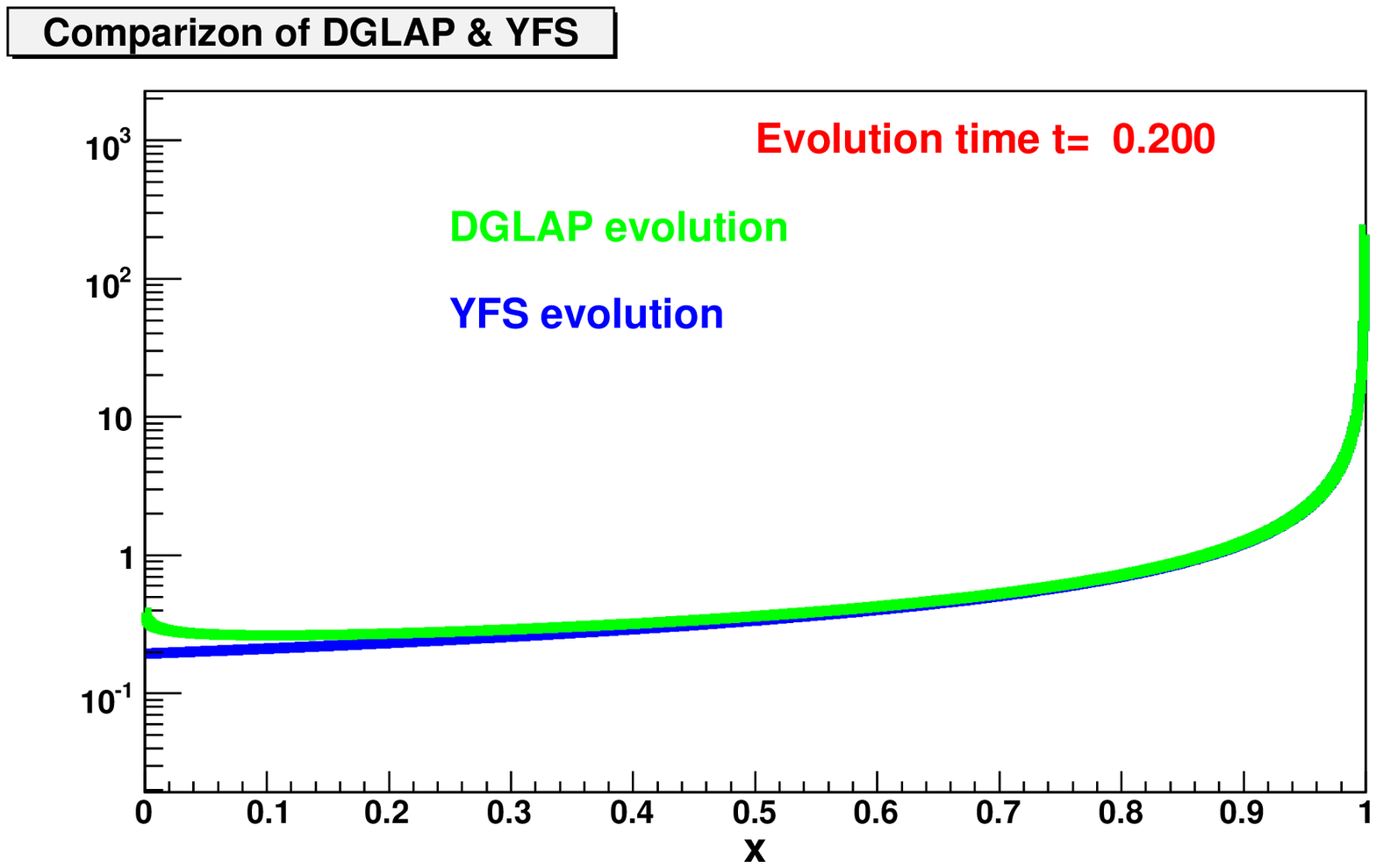, width = 7cm, height=5.6cm}
\epsfig{file=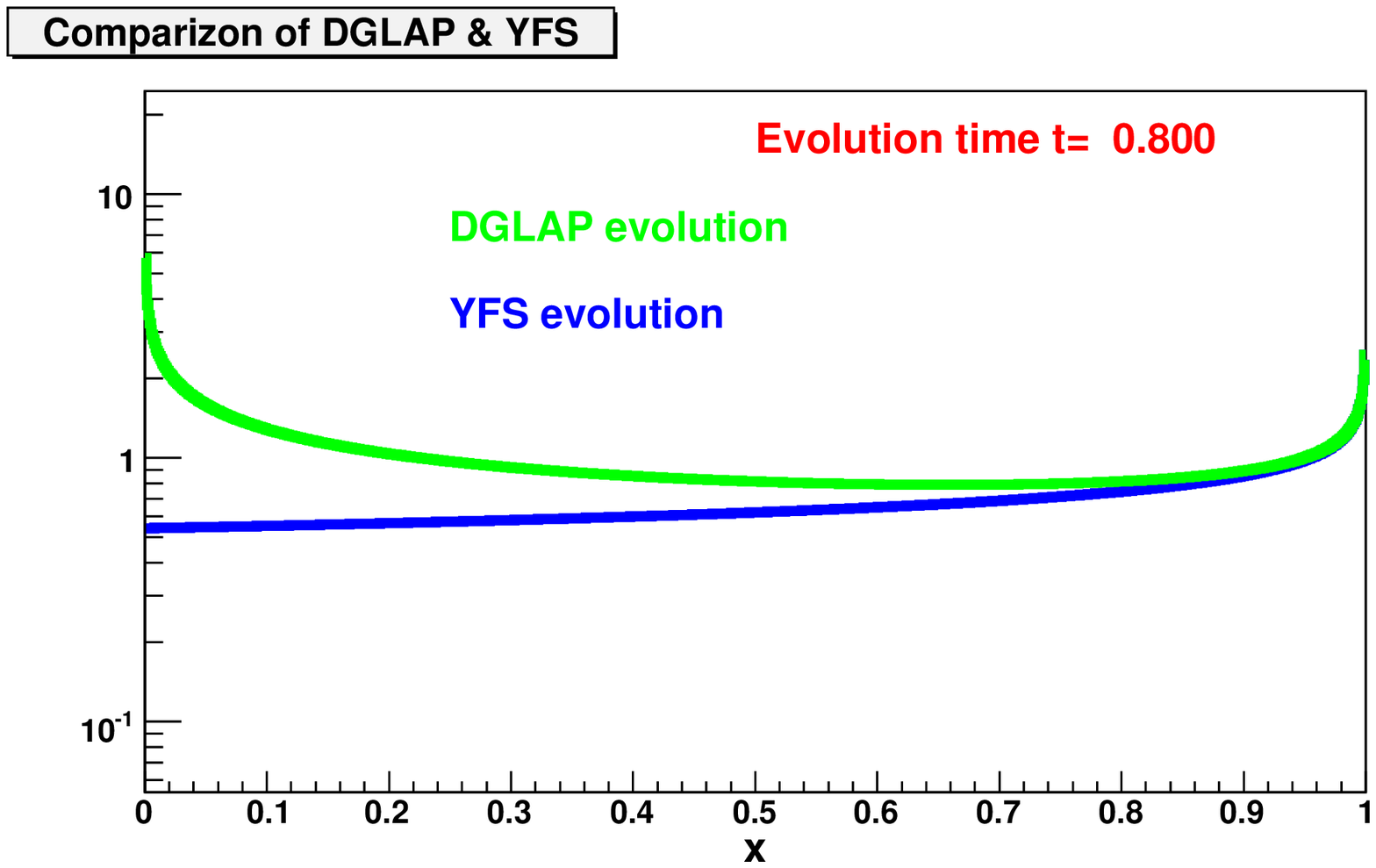, width = 7cm, height=5.6cm}
\caption{Numerical results}
\end{centering}
\end{figure}

The graphs in Figure \ref{YFS_DGLAP} present solutions of \eqref{dyfs_0} and \eqref{DGLAP} with common initial condition  $D(t_0=0,x)=\delta(1-x)$,obtained at chosen evolution times. 

For small evolution times, both distributions are almost identical. The best agreement is in the region $x \approx 1$, where both \eqref{reg_K} and  \eqref{reg2_P} are very similar. In QCD $D(t|x\approx 1)$  represents contribution to the total parton density function coming from very few/ very soft emissions that do not change the initial momentum of the branching gluon. In YFS evolution this region corresponds to $y \approx 0$ in its physical variables meaning that emitted photons are ``reasonably soft'' and do not break the energy conservation. For small $t$ one solution may be approximated by the other.

The discrepancies arise for small $x$, due to differences in singular parts of evolution kernels. They  grow larger with increasing $t$. 

 For $t \rightarrow \infty$, DGLAP solution converges to $\delta(x=0)$ , and YFS has a non-singular asymptotics at $x \rightarrow 0$, $\sim x^{at}$. The right graph in Figure \ref{YFS_DGLAP} presents comparison of solutions for large evolution time. It is visible, that the normalization of $D_{YFS}$ decreases, whereas the normalization of $D_{DGLAP}$ is constant. 

\section*{Acknowledgments}
The project is partly supported by EU grant MTKD-CT-2004-510126.

\section*{References}

\bibliography{moriond}{}

\begin{thebibliography}{1}

\bibitem{yfs:1961}
D.~R. Yennie, S.~Frautschi, and H.~Suura.
\newblock The infrared divergence phenomena and high-energy processes.
\newblock {\em Ann. Phys. (NY)}, 13:379, 1961.

\bibitem{DGLAP}
L.N. Lipatov, {\em Sov. J. Nucl. Phys.} {\bf 20} (1975) 95;\\ V.N. Gribov and
  L.N. Lipatov, {\em Sov. J. Nucl. Phys.} {\bf 15} (1972) 438;\\ G. Altarelli
  and G. Parisi, {\em Nucl. Phys.} {\bf 126} (1977) 298;\\ Yu. L. Dokshitzer,
  {\em Sov. Phys. JETP} {\bf 46} (1977) 64.

\bibitem{Ellis:1991qj}
R.~K. Ellis, W.~James Stirling, and B.~R. Webber.
\newblock Qcd and collider physics.
\newblock {\em Camb. Monogr. Part. Phys. Nucl. Phys. Cosmol.}, 8:1--435, 1996.

\end{thebibliography}

\end{document}